# A new optimization problem in FSO communication system

Mohammad Ali Amirabadi, and Vahid Tabataba Vakili

*Abstract*—According to the physical phenomena of atmospheric channels and wave propagation, performance of wireless communication systems can be optimized by simply adjusting its parameters. This way is more economically favorable than consuming power or using processing techniques. In this paper for the first time an optimization problem is developed on the performance of free-space optical multi-input multi-output (FSO-MIMO) communication system. Also it is the first time that the optimization of FSO is developed under saturated atmospheric turbulences. In order to get closer to the actual results, the effect of pointing error is taken into considerations. Assuming MPSK, DPSK modulation schemes, new closed-form expressions are derived for Bit Error Rate (BER) of the proposed structure. Furthermore, an optimization is developed taking into account the beam width as the variable parameter, and BER as the objective function, there is no constraint in this system. The obtained results can be a useful outcome for FSO-MIMO system designers in order to limit effects of pointing error as well as atmospheric turbulences and thus achieves optimum performance.

*Index Terms*—Free Space Optical Communication, Multi-input Multi-output, Saturate Atmospheric Turbulence, Pointing Error;

## I. INTRODUCTION

Due to the considerable demand for capacity and data rate in the next generation communication systems, communicating over the optical domain, the so called FSO system, with unlimited, unlicensed spectrum, has been proposed as an alternative for conventional wireless systems. FSO system hardware can support multiple Gsps, and typically uses pulse-based modulations, such as on-off-keying (OOK) or pulse position modulation (PPM) [1]. One of the well-known modulations used in FSO systems is subcarrier intensity modulation (SIM), which does not need adaptive threshold detection and is more rugged to the atmospheric turbulences and provides satisfactory performance. This modulation leverages on advances made in signal processing as well as revolution of Radio Frequency (RF) devices such as highly selective filters and stable oscillators, and permits the use of modulation techniques such as phase shift keying (PSK) and quadrature amplitude modulation (QAM) [2].

FSO system has important role in the architecture of hybrid wireless-communications, due to its feasibility in last-mile access, service acceleration, metro network extensions, enterprise connectivity, backup and backhaul links. In the urban environments, the FSO transreceiver should be mounted on high buildings to obtain a Line of Sight (LOS). The availability of a LOS is affected by transreceiver misalignment, which is called pointing error. Also intensity fluctuations caused by atmospheric turbulence, degrades FSO system performance. Mitigate the effects of pointing error and atmospheric turbulence is an important issue in design of FSO system. Which can be done by appropriately adjusting system parameters or using the efforts of multi-input multi-output (MIMO) scheme [3].

Transreceiver misalignment can be caused by winds, thermal expansions, and earthquakes. Under the influence of wind, high-rise buildings sway in three directions of along wind, across wind, and torsional. Transreceiver misalignment is a random process that affects system performance by means of the pointing error [4].

Even at clear weather, FSO system is uncounted with atmospheric turbulence. This effect is like fading in RF system and causes random fluctuations in signal intensity [5]. Following statistical models have been developed to investigate this effect; Exponential-Weibull [6], Generalized Malaga [7], Log-normal [8], Gamma-Gamma [9], and Negative Exponential [10]. Among them Negative Exponential model has high accompany with experimental results for saturated atmospheric turbulence.

Recently, some investigations were developed on the optimization of FSO system. A minimization model for transmitter power and optimization model for divergence angle in a given Bit Error Rate (BER) are developed in [4]. However, it has not provided closed-form expressions. Two optimization models for FSO systems are presented in [3] based on [4], and wavelength is taken as varying parameter. A FSO system in atmospheric turbulence and pointing error is considered in [11], beam width, pointing error variance, and detector size are taken into account; lognormal and gamma-gamma atmospheric turbulences are considered. The BER expression for an intensity-modulation/direct detection (IM/DD) FSO system in strong atmospheric turbulence and pointing error is derived in [12]. [13] assumed IM/DD in the general model of misalignment given in [11]. It did not consider any atmospheric turbulence effects.

In this paper a FSO-MIMO communication system is investigated under the effect of saturated atmospheric turbulence with pointing error. To the best of the authors' knowledge, it is the first time an optimization model is developed over FSO-MIMO systems, also it is the first time that saturated atmospheric turbulence is considered in an FSO optimization problem. Assuming MPSK, DPSK modulation schemes, new closed-form expressions are derived for BER of the proposed structure. Furthermore, an optimization is developed taking into account the beam width as the variable parameter, and BER as the objective function, there is no constraint in this system.

## II. SYSTEM MODEL

Consider a FSO system with N transmit and M receive apertures, where $x$ is the transmitted signal by all transmit apertures over one time slot. After optical-to-electrical conversion, the received electrical signal at $i-th, i = 1,..,M$ receive aperture, becomes as follows:

$$y_i = \eta \sum_{j=1}^{N} I'_{i,j} x + e_i ,  \quad (1)$$

where $e_i$, additive white Gaussian noise (AWGN), is zero mean and $\sigma^2$ variance; $I'_{i,j}$ is irradiance of the link between the $j-th, j = 1,..,N$ transmit and $i-th$ receive aperture, containing the effects of Negative Exponential distributed atmospheric turbulence and pointing error; $\eta$ denotes the optical-electrical conversion efficiency. It is assumed that $\eta = 1$, and $E[|x|^2] = E_x$, where $E[\cdot]$ stands for the expectation. The Equal Gain Combiner (EGC) is used to combine received electrical signals as follows [2]:

$$y = \sum_{i=1}^{M} y_i = \sum_{i=1}^{M} \sum_{j=1}^{N} I'_{i,j} x + \sum_{i=1}^{M} e_i . \quad (2)$$

Because pointing error and atmospheric turbulence affect the LOS, received power is obtained by multiplying the transmitter power ($P_T$), transmitter and receiver telescope gains ($G_T, G_R$), and losses and is given as:

$$P_R = \left(P_T h \sum_{i=1}^{M} \sum_{j=1}^{N} I'_{i,j}\right) \eta_T \eta_R \left(\frac{\lambda}{4\pi d}\right)^2 G_T G_R L_A L_T \quad (3)$$

where $h$ is the random variable indicating pointing error, $\eta_T$ is the optical efficiency of the transmitter and $\eta_R$ is the optical efficiency of the receiver, $\lambda$ is the wavelength, d is the distance between transmitter to receiver, $L_A$ is the atmospheric loss, and $L_T$ is the transmitter pointing loss factor. The term in parentheses is the free-space loss [4]. In this paper it is assumed that the gains and optical efficiencies have unit value, and losses are omitted; therefore only the terms in first parentheses remain, i.e.

$$P_R \approx P_T h \sum_{i=1}^{M} \sum_{j=1}^{N} I'_{i,j} = P_T h I' . \quad (4)$$

At the detector, $P_R$ is converted to electric current $I_R$. The relation between them can be expressed as:

$$I_R = \rho P_R + \rho P_b + I_d + n, \quad (5)$$

where $P_b$ is the received background radiation, $I_d$ is the dark current in the photo-diode, $n$ is the receiver noise, and $\rho$ is the detector responsivity. The effect due to $P_b$ and $I_d$ can usually be compensated with a proper set-up, thus electric current becomes as follows:

$$I_R \approx \rho P_R + n = \rho P_T h I' + n = \rho P_T I + n . \quad (6)$$

For typical photo-emissive and semiconductor junction detectors, the responsivity is described as $\rho = \eta q \lambda / (h_0 c)$, where c is the speed of light, $h_0$ is Plank's constant ($6.626069 \times 10^{-34}$ joule second), $q$ is the electron charge, and $\eta$ is the detector's quantum efficiency, defined as the ratio of the number of emitted electrons to the number of incident photons [3].

Assuming a Gaussian spatial intensity profile of beam waist $w_z$ on the receiver plane at distance z from the transmitter and a circular aperture of radius r, the probability density function (pdf) of h is given by:

$$f_h(h) = \frac{\xi^2}{A_0^{\xi^2}} h^{\xi^2 - 1}; 0 \leq \xi \leq A_0 , \quad (7)$$

where $\xi = w_{z_{eq}}/2\sigma_s$ is the ratio between the equivalent beam radius at the receiver and the pointing error displacement standard deviation at the receiver, $w_{z_{eq}}^2 = w_z^2 \frac{\sqrt{\pi}\mathrm{erf}(\upsilon)}{2\upsilon e^{-\upsilon^2}}$, $\upsilon = \sqrt{\pi}r/\sqrt{2}w_z$, $A_0 = [\mathrm{erf}(\upsilon)]^2$, and $\mathrm{erf}(\cdot)$ is the error function [13].

Considering unit variance Negative Exponential atmospheric turbulence, the pdf of $I_{i,j}$ is given by [14]:

$$f_{I'_{i,j}}(I') = e^{-I'} . \quad (8)$$

Moment Generation Function (MGF) of $I'_{i,j}$, becomes in the following form:

$$M_{I'_{i,j}}(s) = \frac{1}{s+1} . \quad (9)$$

Thus considering independent identically distributed FSO path, the MGF of $I' = \sum_{i=1}^{M} \sum_{j=1}^{N} I'_{i,j}$ becomes as:

$$M_{I'}(s) = \left(\frac{1}{s+1}\right)^{MN}. \quad (10)$$

Therefore, the pdf of $I'$ becomes as:

$$f_{I'}(I') = \frac{I'^{MN-1}}{\Gamma(MN)} e^{-I'} . \quad (11)$$

According that $I = hI'$, the pdf of $I$ becomes equal to:

$$f_I(I) = \int_0^\infty f_{I'}(I') f_h(I/I') dI' = \quad (12)$$
$$\int_0^\infty \frac{\xi^2}{A_0^{\xi^2} \Gamma(MN)} \left(\frac{I}{I'}\right)^{\xi^2 - 1} I'^{MN-1} e^{-I'} dI' .$$

Using [15,Eq.06.05.02.0001.01], the pdf and Cumulative Distribution Function (CDF) of $I$ become equal to:

$$f_I(I) = \frac{\xi^2 \Gamma(MN - \xi^2 + 1)}{A_0^{\xi^2} \Gamma(MN)} I^{\xi^2 - 1} . \quad (13)$$

$$F_I(I) = \frac{\Gamma(MN - \xi^2 + 1)}{A_0^{\xi^2} \Gamma(MN)} I^{\xi^2} . \quad (14)$$

### III. OUTAGE PROBABILITY

Since OOK modulation is used, $x$ is either 0 or $2P_T$ where $P_T$ is the average transmitted optical power. The received electrical SNR and the electrical average SNR, can be defined as [13]:

$$\gamma = \frac{2P_T^2 \rho^2 I^2}{\sigma_n^2}, \gamma_{avg} = \frac{2P_T^2 \rho^2}{\sigma_n^2}. \quad (15)$$

The outage probability denotes the probability that received electrical SNR falls below a threshold SNR, and can be calculated in the following form [13]:

$$Pr(\gamma \leq \gamma_{th}) = F_I\left(\sqrt{\frac{\gamma_{th}}{\gamma_{avg}}}\right) = \frac{\Gamma(MN - \xi^2 + 1)}{A_0^{\xi^2} \Gamma(MN)} \left(\frac{\gamma_{th}}{\mu}\right)^{\frac{\xi^2}{2}} . \quad (16)$$

### IV. BIT ERROR RATE

The average BER, can be derived as

$$P_b(e) = \int_0^\infty f_I(I) P_b(e|I) \, dI , \quad (17)$$

where $P_b(e|I)$ is the BER for channel conditioned on $I$.
For MPSK, the instantaneous BER is given by [2]:

$$P_b(e|I) = \frac{\zeta_M}{2} \sum_{p=1}^{\tau_M} erfc\left(\frac{a_p P_T \rho I}{\sigma_n}\right) , \quad (18)$$

where $erfc(\cdot)$ is the complimentary error function, $\zeta_M = \frac{2}{\max(\log_2 M, 2)}$, $a_p = \sqrt{2} \sin\frac{(2p-1)\pi}{M}$, and $\tau_M = \max\left(\frac{M}{4}, 1\right)$ are the modulation dependent parameters of an MPSK constellation containing M-points.

Substituting (13) and (18) into (17), and using [15,Eq.06.27.21.0132] the average BER will be:

$$P_b(e) = \quad (19)$$
$$\sum_{p=1}^{\tau_M} \frac{\zeta_M}{2} \frac{\xi^2 \Gamma(MN - \xi^2 + 1)}{A_0^{\xi^2} \Gamma(MN)} \left(\int_0^\infty erfc\left(\frac{a_p P_T \rho I}{\sigma_n}\right) I^{\xi^2 - 1} dI\right) =$$
$$\sum_{p=1}^{\tau_M} \frac{\zeta_M}{2\sqrt{\pi}} \frac{\Gamma(MN - \xi^2 + 1) \Gamma\left(\frac{\xi^2 + 1}{2}\right)}{\Gamma(MN)} \left(\frac{2}{A_0^2 a_p^2 \gamma_{avg}}\right)^{\frac{\xi^2}{2}} .$$

Using [15,Eq.06.05.20.0001.01], differentiate of BER will be:

$$\frac{dP_b(e)}{d\xi} = -2\psi(MN - \xi^2 + 1) + \psi\left(\frac{\xi^2 + 1}{2}\right) - \quad (20)$$

$$ln\left(\frac{A_0^2 a_p^2 \gamma_{avg}}{2}\right) .$$

For DPSK, the instantaneous BER is given by [17]:

$$P_b(e|I) = \frac{1}{2} e^{-\frac{2P_T^2 \rho^2 I^2}{\sigma_n^2}} . \quad (21)$$

Substituting (13) and (21) into (17), and using [15,Eq.06.27.21.0132] the average BER will be:

$$P_b(e) = \quad (22)$$
$$\frac{\xi^2 \Gamma(MN - \xi^2 + 1)}{A_0^{\xi^2} 2\Gamma(MN)} \left(\int_0^\infty e^{-2\left(\frac{P_T \rho I}{\sigma_n}\right)^2} I^{\xi^2 - 1} dI\right) =$$
$$\frac{\xi^2 \Gamma(MN - \xi^2 + 1) \Gamma\left(\frac{\xi^2}{2}\right)}{2\sqrt{2} \, \Gamma(MN)} \left(\frac{1}{A_0^2 \gamma_{avg}}\right)^{\frac{\xi^2}{4}} .$$

Using [15,Eq.06.05.20.0001.01], differentiate of BER becomes equal to:

$$\frac{dP_b(e)}{d\xi} = \xi^2 \left(\frac{1}{2} \psi\left(\frac{\xi^2}{2}\right) - \psi(MN - \xi^2 + 1)\right) - \quad (23)$$
$$\frac{1}{2} ln\left(\frac{\sqrt{2} P_T \rho A_0}{\sigma_n}\right) + 1 .$$

The obtained BER for DPSK and MPSK, considering beam width as a variable parameter can be minimized by finding the root of $\frac{dP_b(e)}{d\xi} = 0$, respectively in (20), and (23). MATLAB solve[.] command can easily solve them.

### V. RESULTS AND DISCUSSIONS

In Fig. 2, Bit Error Rate of the proposed FSO-MIMO structure is plotted as a function of average SNR and $\xi$, for BPSK modulation, when number of transmitter and receiver aperture is $M = N = 6$. As can be seen, BER reduces while increasing $\xi$, this reduction continues till reaching a specific, e.g. at $\gamma_{avg} = 0dB$, this occurs about $\xi = 5.5$. This specific $\xi$ changes at different $\gamma_{avg}$. However, it increases while increasing $\gamma_{avg}$. The performance of FSO system can be optimized without additional processing, and computation with adjusting system parameters; Fig. 2 indicates this argument in the term of beam width; because $\xi$ is is the ratio of beam width to the jitter variance. One can assume constant jitter variance and adjust beam width accordingly.

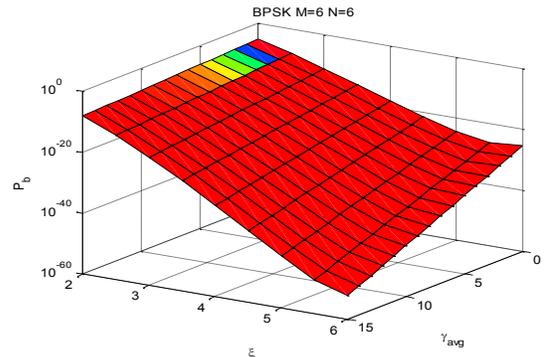

Fig. 2. Bit Error Rate of the proposed FSO-MIMO structure as a function of average SNR and $\xi$, for BPSK modulation, when number of transmitter and receiver aperture is $M = N = 6$;

In Fig. 4, Outage Probability of the proposed FSO-MIMO structure is plotted as a function of normalized SNR and $\xi$, when number of transmitter and receiver aperture is $M = N = 6$. It can be seen that the reduction in BER is smother than Outage Probability while increasing $\xi$.

In Fig. 5, Bit Error Rate of the proposed FSO-MIMO structure as a function of average SNR, for DBPSK, and BPSK modulations, when number of transmitter and receiver aperture is $M = N = 6$; as can be seen, performance of BPSK is better than DBPSK, but differential modulations such as DBPSK, are less sensitive to noise and interference and do no require complex processing .

## VI. Conclusion

In this paper a FSO-MIMO communication system is considered under the effects of pointing error and saturated atmospheric turbulence. Assuming MPSK, DPSK modulation schemes, new closed-form expressions are derived for BER of the proposed structure. Furthermore, in order to mitigate effects of pointing error and saturated atmospheric turbulence, a minimization problem is developed in which BER is the objective function and beam width is the variable parameter, there is no constraint assumed in this problem. The obtained results can be useful outcome for FSO-MIMO system designers in order to achieve the optimum performance by adjusting natural system parameters, without additional processing complexity and latency.

Results indicate that BER reduces while increasing beam width, this reduction continues till reaching a specific beam width, which is different at various average SNRs. It is shown that performance of FSO system can be optimized without additional processing, computation or complexity by simply adjusting system parameters such as beam width. This way is more economically favorable than consuming power or using processing techniques.

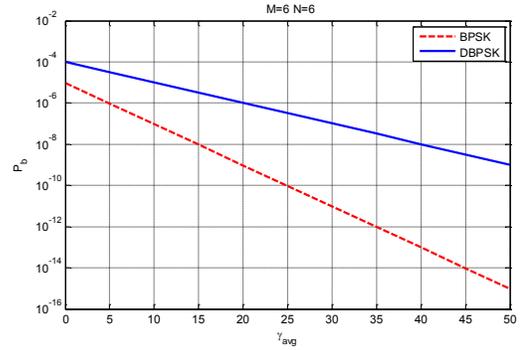

Fig. 5. Bit Error Rate of the proposed FSO-MIMO structure as a function of average SNR, for DBPSK, and BPSK modulations, when number of transmitter and receiver aperture is $M = N = 6$;

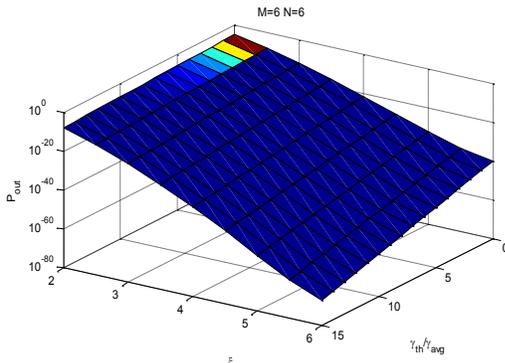

Fig. 4. Outage Probability of the proposed FSO-MIMO structure as a function of normalized SNR and $\xi$, when number of transmitter and receiver aperture is $M = N = 6$;